%% file: mult_pion_2016.tex
\documentclass[a4paper,manyauthors,nocleardouble,COMPASS]{cernepprep}

\graphicspath{  {./figures/} }

\usepackage{bm}
\usepackage{color}
\input myunits.sty

\usepackage{amsmath}
\usepackage{color}
\usepackage[T1]{fontenc}
\usepackage[english]{babel}
\usepackage{amsmath}
\usepackage{bm}
\usepackage{amsfonts}
\usepackage{mathrsfs}
\usepackage{amssymb}
\usepackage[ansinew]{inputenc}
\usepackage[font=small,format=plain,labelfont=bf,up]{caption}
\usepackage{hyperref}
\usepackage{cite}
\usepackage[dvipsnames]{xcolor}
\usepackage{pdfpages}
\usepackage{multirow}
\usepackage{array}
\usepackage{subcaption}
\usepackage{color}
 
\makeatletter

\DeclareSymbolFont{letters}     {OML}{cmm}{m}{it}
\DeclareSymbolFont{symbols}     {OMS}{cmsy}{m}{n}
\DeclareSymbolFont{largesymbols}{OMX}{cmex}{m}{n}

\begin{document}

\begin{titlepage}

\EPnumber{2016--095}
\EPdate{\raisebox{0pt}[0pt][0pt]{\raisebox{-1.9ex}{\shortstack[l]{10 April 2016\\
\small rev. 28 April 2016}}}}

\title{Multiplicities of charged pions and unidentified charged hadrons from deep-inelastic scattering of muons off an isoscalar target}

\Collaboration{The COMPASS Collaboration}
\ShortAuthor{The COMPASS Collaboration}

\begin{abstract}

Multiplicities of charged pions and unidentified hadrons
produced in deep-inelastic scattering
were measured in bins of the Bjorken scaling variable $x$, 
the relative virtual-photon energy $y$
and the relative hadron energy $z$. Data were obtained by the COMPASS 
Collaboration using a $160\,\GeV$ muon beam and an isoscalar target ($^6$LiD).
They cover the kinematic domain in the photon virtuality 
$Q^2 > 1\,(\GeV/c)^2$, $0.004 < x < 0.4$, $0.2 < z < 0.85$ 
and $0.1 < y < 0.7$. 
In addition, a leading-order pQCD analysis was performed using the pion 
multiplicity results to extract quark fragmentation functions.
\end{abstract}

\vspace*{60pt}

\vfill
\Submitted{(to be submitted to Phys.\ Lett.\ B)}

\end{titlepage}

{\pagestyle{empty} \input{Authors2016.tex}}
\newpage
\setcounter{page}{1}
\parindent=0em

\section{Introduction}
\vspace{0.2cm}

Hadron production in semi-inclusive measurements of deep-inelastic 
lepton--nucleon scattering (SIDIS) is one of the 
most powerful tools to investigate the structure and formation of hadrons. 
Within the standard framework of leading-twist perturbative QCD (pQCD), 
factorisation theorems~\cite{collins} allow one
to write the SIDIS cross section as a convolution of hard scattering cross sections,
which are calculable in pQCD, with non-perturbative Parton Distribution Functions (PDFs) and
Fragmentation Functions (FFs). The
PDFs account for the partonic structure of hadrons 
in the initial state. The FFs encode the details about the hadronisation 
mechanism {that}
describes the transition from final-state partons into colour-neutral 
hadrons. Both types of functions are believed to be universal, 
i.e. process independent, and can be 
interpreted in leading-order (LO) as number densities. 
While PDFs have 
been studied in detail for several decades and are hence known with good 
precision,   
new accurate measurements are necessary to constrain the FFs.
In what follows, 
we will restrict ourselves to transverse-momentum-integrated PDFs and FFs.

The universality of FFs allows their determination from different 
high-energy processes, which provide 
complementary information on the hadronisation mechanism and
cover complementary kinematic ranges. The study of hadron production in the 
electron--positron annihilation process 
is particularly well suited because its cross section
has no dependence on parton densities and gives direct access to FFs.
Measurements cover a wide range in the characteristic hard scale, from 
$10\,\GeV$ for recent data from BELLE~\cite{Leitgab:2013qh} and 
BABAR~\cite{babar} to $100\,\GeV$ at the Z boson mass at LEP~\cite{LEP} and 
SLAC~\cite{SLAC}, at which
only the singlet combination of FFs is accessible. In spite of the high 
precision, these e$^+$e$^-$ data cannot be used to disentangle quarks from
anti-quarks, as they only access the sum of quark and antiquark FFs 
and hence do not allow for a full flavour separation. 
Hadron multiplicity
data from SIDIS provide charge and full flavour
separation of fragmentation functions. The SIDIS data from fixed target experiments 
explore characteristic hard scales down to $1\,\GeV$. The large kinematic
range spanned by the mentioned reactions allows one to study QCD
scaling violations, which also
constrains the gluon FF. The latter is indirectly probed by 
hadron-hadron collisions, e.g.~at RHIC~\cite{RHIC}, via single-inclusive 
hadron production at high transverse momentum~\cite{deFlorian:2007aj}.

The present Paper reports on COMPASS measurements of multiplicities of charged pions and unidentified charged hadrons in a kinematic range
that is larger than the one covered by HERMES~\cite{HERMES} and similar to 
the one covered by EMC~\cite{EMC}.

The process ${\rm l}\, {\rm N} \rightarrow {\rm l}'\,{\rm h}\, {\rm X}$ is described 
by the negative square of the four-momentum transfer $Q^{2}= -q^{2}$, 
the Bjorken variable $x = -q^{2}/(2P\cdot q)$ and the fraction of the 
virtual-photon energy that is carried by the final-state hadron, 
$z = (P\cdot p_{\rm h})/(P\cdot q)$. 
Here, $q=k-k'$, $P$ and $p_{\rm h}$ denote the four-momenta of the virtual photon, the 
nucleon N and the observed hadron h respectively, with $k$ ($k'$) the four 
momentum of the incident (scattered) lepton.
Additional variables used are the lepton energy fraction carried by the 
virtual photon, $y=(P\cdot q)/(P\cdot k)$, and the invariant mass 
of the final hadronic system, $W = \sqrt{(P + q)^{2}}$. 
In order to study the hadronisation mechanism in SIDIS, the relevant observable
is the differential multiplicity for 
hadrons of a specific type h, which is defined as the differential cross section for 
hadron production 
normalised to the differential inclusive DIS cross section:
\begin{equation}
\frac{{\rm d}M^{\rm h}(x,z,Q^2)}{{\rm d}z}=
\frac{{\rm d}^3\sigma^{\rm h}(x,z,Q^2)/{\rm d}x{\rm d}Q^2{\rm d}z}{{\rm d}^2\sigma^{\rm DIS}(x,Q^2)/{\rm d}x{\rm d}Q^2}.
\label{MulDef}
\end{equation}
Interpreted in pQCD, the cross sections on the {right-hand side}
are expressed in terms of PDFs and FFs and read at leading-order (LO)

\begin{equation}
\frac{{\rm d}^2\sigma^{\rm DIS}}{{\rm d}x{\rm d}Q^2}=C(x,Q^2)\sum_qe_q^2q(x,Q^2), \qquad
\frac{{\rm d}^3\sigma^{\rm h}}{{\rm d}x{\rm d}Q^2{\rm d}z}=C(x,Q^2)\sum_qe_q^2q(x,Q^2)D_q^{\rm h}(z,Q^2).
\label{Sigma}
\end{equation}
Here, $q(x,Q^2)$ is the quark PDF for the flavour $q$, $D_q^{\rm h}(z,Q^2)$ 
the quark-to-hadron FF, 
$C(x,Q^2)={2\pi\alpha^2(1+(1-y)^2)}/{Q^4}$ and $\alpha$ the fine structure 
constant. In LO, 
$D_{\rm q}^{\rm h}$ denotes 
the number density of hadrons h produced in the hadronisation of partons {of species} q.

\section{The COMPASS experiment}
\vspace{0.2cm}
In this Section a short description of the experimental set-up is given, while
a more detailed description can be found in Ref.\,\citen{Abbon:2007pq}.
The measurement was performed in 2006 with the naturally polarised muon beam 
of the CERN SPS using positive muons of $160\,\GeV/c$. The beam momentum had 
a spread of 5\%. The intensity was $4\times 10^7\,\s^{-1}$ with spills of 
$4.8\,\s$ and a cycle time of $16.8\,\s$. The momentum of each incoming muon 
was measured at the end of the beam line with a precision of 0.3\%. 
Before the target, the trajectory of each incoming muon was measured in a 
set of silicon and scintillating fibre
detectors with a precision of $0.2\,\mrad$. 
The muons were impinging on a longitudinally polarised solid-state target 
positioned inside a large aperture solenoid. The target consisted of three 
cells, which were located along
the beam one after the other and filled with $^6$LiD immersed in a liquid 
$^3$He/$^4$He mixture. The admixtures of H, $^3$He and $^7$Li in the isoscalar 
target lead to an effective
excess of neutrons of about 0.2\%. 
{The direction of the polarisation in the $60\,\Cm$ long 
middle cell was opposite to that in the two $30\,\Cm$ long outer cells.}
In the analysis,
the data are averaged over the target polarisation for the determination 
of multiplicities. 

The two-stage COMPASS spectrometer was designed to reconstruct scattered muons
and produced hadrons in a wide range of angle and momentum. Particle tracking 
was performed by a variety of tracking detectors before and after the two 
spectrometer
magnets. The direction of the reconstructed tracks at the interaction point is
determined with a precision of $0.2\,\mrad$, and  the momentum resolution is
1.2\% in the first spectrometer stage and 0.5\% in the second. Muons are
identified downstream of hadron absorbers. A ring imaging Cherenkov 
counter (RICH) in the first stage is used for pion, kaon and proton 
separation~\cite{RICH}.
It was filled with a C$_4$F$_{10}$ radiator leading to thresholds for
pion, kaon and proton detection of about $2.9\,\GeV/c$, $9\,\GeV/c$ and 
$18\,\GeV/c$ respectively. {In the central
part, photon detection was performed using} multi-anode photomultiplier tubes
that yielded high photodetection efficiency and a fast response in the
high rate environment. {In the outer part,} multi-wire proportional chambers 
with CsI cathodes were used to detect the UV Cerenkov photons. The trigger,
based on pairs of hodoscopes,
selected scattered muons above a minimum scattering angle.

\section{Data analysis}
The data analysis includes event selection, particle 
identification (PID), acceptance correction as well as corrections for radiative effects and
diffractive vector meson production.
Differential multiplicities are determined in 3-dimensional 
($x$, $y$, $z$) bins from 
the acceptance-corrected hadron yields $N^{\rm h}$
normalised by the number of DIS events, $N^{\rm DIS}$: 
\begin{equation}
\label{ExpMul}
\frac{\text{d}M^{\rm h}(x,y,z)}{\text{d}z} = 
\frac{1}{N^{\rm DIS}(x,y)}\frac{\text{d}N^{\rm h}(x,y,z)}{\text{d}z} \frac{1}{A(x,y,z)}\,.
\end{equation}
The acceptance correction factor $A$ takes into account the limited geometric 
and kinematic acceptance of the spectrometer and the efficiency of event 
reconstruction.
The choice of the $z$ and $x$ variables is natural because
multiplicities depend mostly on these variables. 
Because of the strong correlation between $x$ and $Q^2$ in the 
COMPASS fixed-target kinematics, it appears {more appropriate
to use $y$ instead of $Q^2$ as the third variable.}

\subsection{Event and hadron selection}
The present analysis is based on events with inclusive triggers that use 
only information on the scattered muons.
Selected events are required to have a reconstructed 
interaction vertex associated to an incident and a scattered muon track. This 
vertex has to lie inside the fiducial target volume. The 
incident muon energy is constrained to the interval $[140, 180]\,\GeV$. 
Events are accepted if $Q^{2} > 1\,(\GeV/c)^2$, $0.004<x<0.4$ and  $W>5\,\GeV/c^2$. 
These requirements select the deep-inelastic 
scattering regime and exclude the nucleon resonance region. The relative 
virtual-photon energy is constrained to the range $0.1 < y < 0.7$ to exclude kinematic regions 
where the momentum resolution degrades and radiative effects are most 
pronounced. The 
number of inclusive DIS events selected for this analysis is 
13~$\times~10^{6}$, corresponding to an integrated luminosity of 
0.54~fb$^{-1}$.

For a selected DIS event,
all reconstructed tracks are considered. Hadron tracks must 
be detected in tracking detectors placed before and after the magnet in the first 
stage of the spectrometer.
The fraction of the virtual-photon energy transferred to a final-state hadron 
is constrained to $0.2 \le z \le 0.85$, whereby for an unidentified charged 
hadron the pion mass is assumed. The lower limit avoids the contamination 
from target remnant fragmentation, while the upper one excludes muons wrongly 
identified as hadrons, and it also excludes the region with large diffractive 
contributions. Further constraints on momentum and 
polar angle of the hadrons as well as on $y$ are discussed below.

The corrections for higher-order QED effects are applied on an
event-by-event basis taking into account the target composition. 
For $N^{\rm DIS}(x,y)$ they are computed according to the scheme described in
Ref.~\citen{bardin}. For $\text{d}N^{\rm h}(x,y,z)$ the elastic and quasielastic  radiative
tails are subtracted from the correction. The $z$-integrated radiative 
correction factor for the multiplicities is always below 5\%.
A possible $z$ dependence of radiative corrections
is neglected. 

\subsection{Hadron identification using the RICH detector}
\vspace{0.2cm}
Particle identification (PID) is performed using the RICH detector~\cite{Abbon:2011zza}.
The  identification procedure relies on a likelihood function, which is  based
upon the 
number and distribution of photons that are
detected in the RICH detector and associated 
to a charged particle trajectory. The likelihood values are calculated by comparing 
the measured photo-electron pattern with the one expected
for different mass hypotheses ($\pi$, K, p), taking the distribution of 
background photons into account. The mass is assigned to the detected hadron 
choosing the hypothesis with the maximum likelihood. In order to improve 
the separation between the different mass hypotheses and thus the sample 
purity,  constraints are imposed on
the ratios of the maximum over the other likelihood values.

\begin{figure}[htbp]
\centering
\includegraphics[width=.55\textwidth]{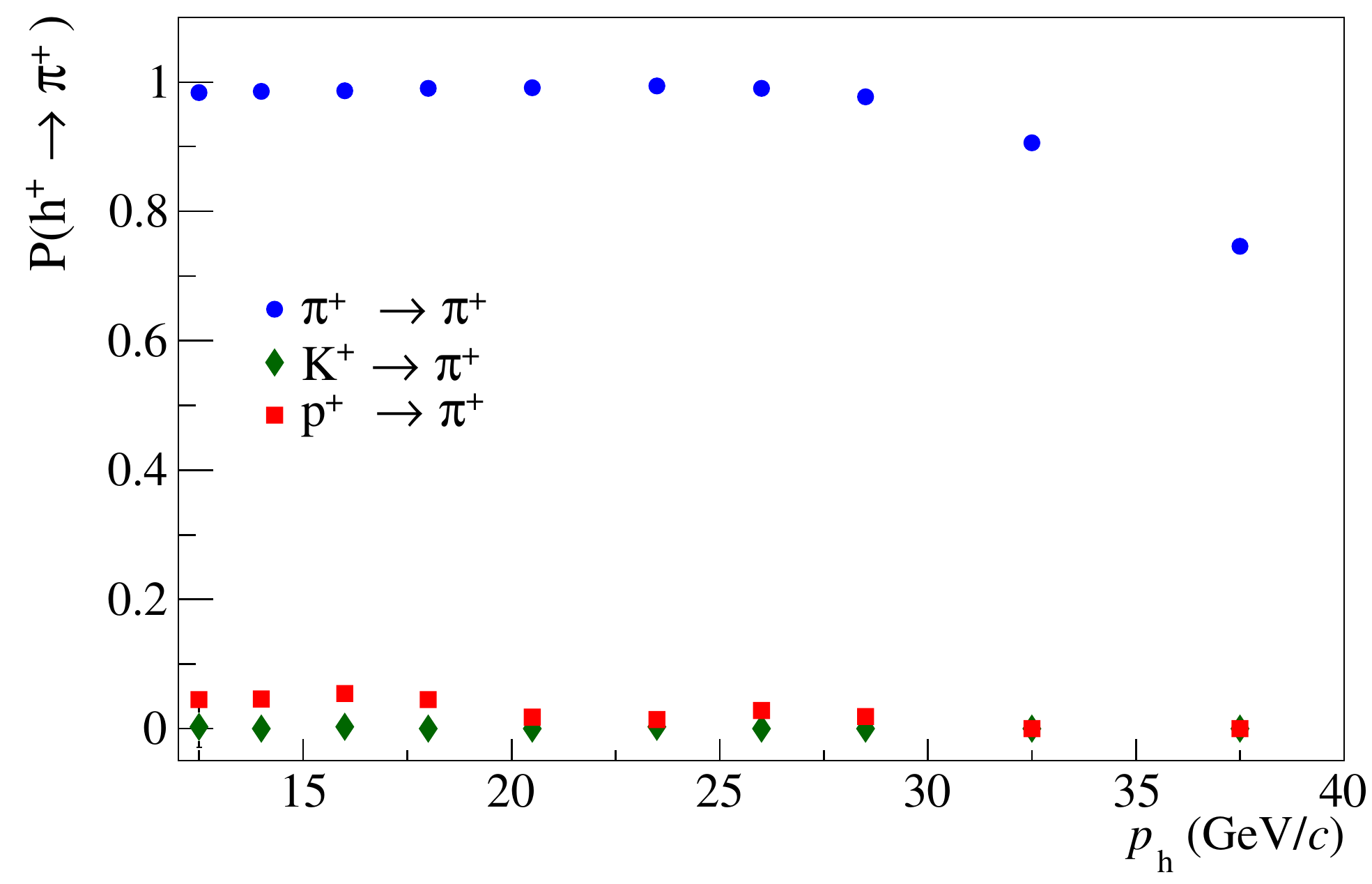}
\caption{Probabilities of RICH identification of $\pi^+$, K$^+$ and p as a 
$\pi^+$ versus momentum for the smaller $\theta$ bin $10\,\mrad <\theta<40\,\mrad$. 
Statistical uncertainties are lower than the size of the symbols.}
\label{RichEff}
\end{figure}

The purity of the identified hadron samples depends on the probabilities of 
correct identification and misidentification. The true hadron 
yields  $N_{\rm true}$ are obtained by applying an unfolding algorithm 
to the measured hadron yields $N_{\rm meas}$:
\begin{equation}
N^{i}_{\rm true} = \sum_{j} (\text{P}^{-1})_{ij} \cdot N^{j}_{\rm meas}\,.
\label{Relation1}
\end{equation}
The RICH PID matrix $\rm{P}$  contains as diagonal elements  the efficiencies and 
as off-diagonal 
elements the misidentification probabilities. 
The elements of this $3 \times 3$ matrix are constrained by 
$\sum_{j} {\rm P}_{ij} \le 1$, where $i,j=\{ \pi, K, {\rm p}\}$. 
They  are determined from real data using samples of $\pi$, K or p 
originating from the decay of K$^0_{\rm S}$, $\phi$ or $\Lambda$ into two charged 
particles.
{The dependence of the RICH performance on the particle 
momentum $p_{\rm h}$ and polar angle $\theta$ at the RICH entrance is taken
into account by determining the RICH PID matrix in 2-dimensional 
bins of these variables.}
The $p_{\rm h}$ dependence accounts for effects arising from momentum 
thresholds {for particle identification}
and from saturation at high momentum. The $\theta$ dependence accounts for 
{varying} 
occupancy and background level in the RICH photon detectors. 
The polar angle is selected in the range  $10\,\mrad < \theta < 120\,\mrad$, where 
the efficiencies are high and  precisely measured. 
The $\theta$ dependence of P$_{ij}$ is relatively weak, so that two bins are
sufficient in the analysis.   
{In order to achieve good pion--kaon 
separation and high particle identification probabilities for kaons and pions,
momenta between $12\,\GeV/c$ and $ 40\,\GeV/c$ are used.
In this range}, the momentum dependence of the probabilities of $\pi^+$, K$^+$ and p to be identified as 
$\pi^+$ is shown in Fig.~\ref{RichEff} for the lower $\theta$ 
{bin in 10 momentum bins}.  
Pions are identified with 98\% efficiency  up to 30 GeV/c, where the  
efficiency starts to decrease. 
The probability to misidentify kaons and protons as pions is below 
2\% and 6\%, respectively, over all the selected momentum range. 
Similar values are obtained for $\pi^-$. 
The number $N^{\pi^\pm}_{\rm true}$ of identified pions 
available for the analysis after all 
PID cuts is 3.4~$\times~10^{6}$. 
Without likelihood cuts, the number of charged 
hadrons is 4.6~$\times~10^{6}$.

\subsection{Acceptance correction}\label{sec:acceptance}
The raw multiplicities must be corrected for the geometric 
and 
kinematic acceptances of the experimental set-up as well as for detector
inefficiencies, resolutions and bin migration.
The full correction is evaluated using a Monte Carlo (MC) simulation of the 
muon--nucleon deep-inelastic scattering process. Events are generated with 
the LEPTO \cite{Ingelman:1996mq} generator, where the parton hadronisation 
mechanism is simulated using the JETSET package \cite{Sjostrand:1995iq} 
with the tuning from Ref.~\citen{Adolph:2012vj}. The spectrometer is 
simulated using the GEANT3 toolkits~\cite{Brun:1985ps}, and the 
MC data are reconstructed with the same software as the experimental 
data~\cite{Abbon:2007pq}. Secondary hadron interactions are
simulated using the FLUKA package~\cite{FLUKA}. 
The kinematic distributions of the experimental data 
are fairly well reproduced by the MC simulation.

In order not to introduce a 
strong dependence on the physics generator used in the simulation, 
the extraction of hadron multiplicities is performed in narrow kinematic bins
of $x$, $y$ and $z$. In each ($x$, $y$, $z$) bin,
the acceptance is calculated from the ratio of reconstructed 
and generated multiplicities according to
\begin{equation}
     A^{\rm h}(x ,y ,z) 
 = \frac { {\rm d} N ^{\text{h}} _{\text{rec}}(x,y,z) / N ^{\text{DIS}} _{\text{rec}}(x,y) } {{\rm d} N^{\text{h}}_{\text{gen}}(x,y,z) /N^{\text{DIS}}_{\text{gen}}(x,y)}.
\label{accep}
 \end{equation}
The generated kinematic variables are used for the generated particles, while 
the reconstructed kinematic variables are used for the reconstructed particles.
After reconstruction, all particles are subject 
to the same kinematic and geometric selection criteria as the data,
while the generated ones are subject to 
kinematic requirements only. At this place, the correction for possible
misidentification of electrons as pions is included in the acceptance
correction.  
The average value of the acceptance is about 70\% for 
$y < 0.3$ and about 50\% for $y > 0.3$. 
The acceptance is almost flat in $z$ and $x$, except at high $y$ and low $x$, {and always larger than 40\%.}

With the hadron momentum cut, the $y$ range is
more limited for the hadron sample than for the DIS sample. 
Thus for each bin in $z$, the 
$y$ range is restricted to the kinematic region accessible with hadron momenta
between $12\,\GeV/c$ and $40\,\GeV/c$.

\subsection{Vector meson correction}
\begin{figure}[htbp]
\centering
\includegraphics[width=0.7\textwidth]{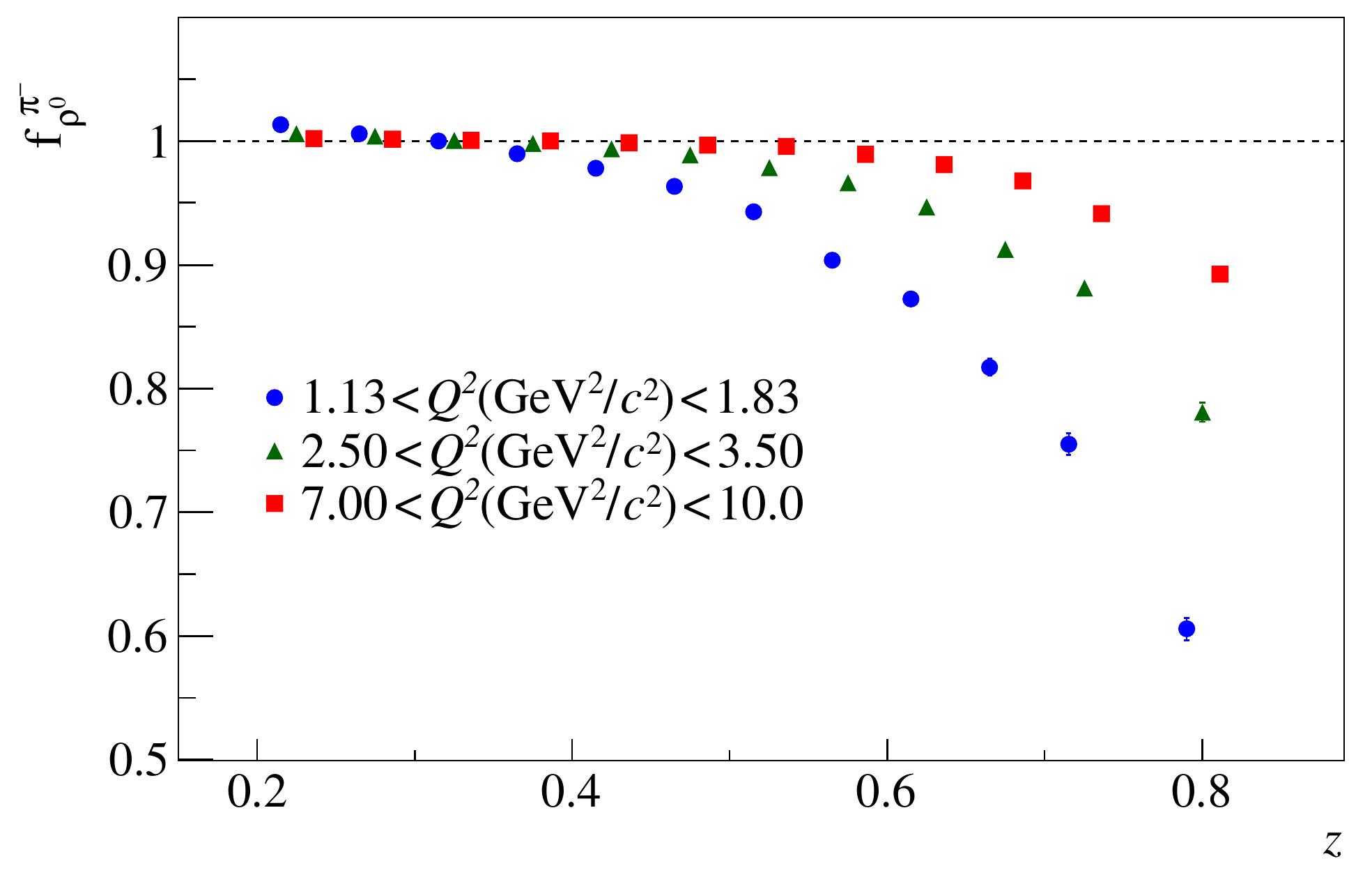}
\caption{Correction due to diffractive $\rho^0$ contamination, shown for negative-pion 
multiplicities as a function of $z$ for three $Q^2$ bins}
\label{fig:rho_contribution}
\end{figure}

A fraction of the mesons measured in SIDIS originates from diffractive 
production of vector mesons, which 
subsequently decay 
into lighter hadrons. This fraction can be considered as a higher-twist 
contribution to the SIDIS cross section~\cite{HERMES}. It cannot be described 
by the QCD parton model with the independent-fragmentation mechanism, which
is encoded in the FFs. Moreover, fragmentation functions extracted from data
including this fraction would be biased, which violates in particular
the universality principle of the model. Therefore, the fraction of final-state
hadrons originating from diffractive $\rho^0$ decay is estimated. Our 
evaluation is based on two MC simulations, one using the LEPTO event 
generator simulating SIDIS free of diffractive contributions 
(see Section 3.3), 
and the other one using the HEPGEN~\cite{hepgen} generator 
simulating diffractive $\rho^0$ 
production. Further channels, which are characterised
by smaller cross sections, are not 
taken into account. 
Events with diffractive 
dissociation of the target nucleon are also simulated and represent about 
25\% of those with the nucleon staying intact. 
The simulation of these events includes nuclear effects,
i.e.\ coherent production and nuclear 
absorption as described in Ref.\,\citen{hepgen}. 
A correction factor for the multiplicities 
is calculated taking into account the diffractive contribution to pion 
(hadron) and DIS yields. As an example, the $z$ dependence of
the correction factor $f^{\pi}_{\rho^0}$ for $\pi^+$ multiplicities
is shown in Fig.\,\ref{fig:rho_contribution} 
for three $Q^2$ bins. The correction varies between 1.02 and 0.55 for pions,
with the largest value appearing {at small $Q^2$
and high $z$, whereby the latter region is characterised
by very small multiplicities.}

\subsection{Systematic uncertainties}
The main contributions to the systematic uncertainties arise from the uncertainties
on the determination of the  acceptance, of the RICH performance and of the 
diffractive $\rho ^0$ contribution.
The uncertainty on the acceptance calculation is evaluated by two 
different methods: first, by varying in the MC the PDF set used and the JETSET parameters related to 
the hadronisation mechanism; secondly, by determining the acceptance in 
a different dimensional space adding additional variables. 
The validity of the correction for electron contamination is
confirmed by comparing the simulated and measured electron distributions 
for momenta below $8\,\GeV/c$, where electrons are identified using the RICH. An uncertainty
of 4\% is found that includes the uncertainty of the electron contamination.

In order to estimate the uncertainty linked to the RICH identification and unfolding 
procedure, 
different RICH matrices are built by varying the matrix elements within their 
statistical 
uncertainties. The differences between the resulting multiplicities 
and the original ones yields an 
estimate of the uncertainty, which is below 1\% for $z<0.4$ and reach  
2\% at high $z$.
No time dependence is observed when comparing the results obtained from 
the data taken in six different weeks.

The cross section for exclusive production of $\rho ^0$  calculated in HEPGEN is 
normalised to the phenomenological  model of Ref.~\citen{GK}. 
The theoretical uncertainty on the 
predicted cross section close to COMPASS kinematics amounts to about 30\%. This 
results in an uncertainty on the diffractive $\rho^0$ correction 
factor, which amounts to at most 30\% and depends on the kinematic range.

Nuclear effects may be caused by the presence of $^3$He/$^4$He and 
$^6$Li in the target. A detailed study of such effects was 
previously performed by the EMC~\cite{EMC} in a 
similar kinematic range for carbon, copper and tin. A 
$z$-dependent decrease of 5\% was observed for the multiplicities of
copper compared to the ones of deuterium. While 
the effect was larger for tin, no such effect was found for carbon,
so that possible nuclear effects in the present experiment are expected 
to be very small and are hence neglected.

All contributions to the systematic uncertainties are added in quadrature and 
yield the total systematic uncertainty shown as bands in Figs.\,\ref{res3} 
and\,\ref{res4}, which
varies between 5\% and 10\%. Note that not all systematic uncertainties
are correlated from bin to bin. It was estimated that, when considering 
quadratic summation, about 80\,\% of 
the total systematic uncertainty is correlated from bin to bin. In this case.
the remaining 60\% is uncorrelated and is treated together
with the statistical uncertainties in the fits discussed in 
Section~\ref{sec:fragmentation}.

\newcommand{\fav}{\text{fav}}
\newcommand{\unf}{\text{unf}}
\newcommand{\Dfav}{D_{\fav}}
\newcommand{\Dunf}{D_{\unf}}
\newcommand{\Dg}{D_{\g}}
\newcommand{\ubar}{\overline{\text{u}}}
\newcommand{\dbar}{\overline{\text{d}}}
\newcommand{\sbar}{\overline{\text{s}}}
\newcommand{\normDIS}{5(u + d + \ubar + \dbar) + 2(s + \sbar)}

\section{Results for pion and unidentified hadron multiplicities}
The multiplicities presented in the following figures are all corrected for the
 diffractive $\rho^0$ contribution.  The numerical values are 
available on HepData\cite{HEPDATA} for multiplicities  
with and without this correction. 
The separate correction factors for DIS and pion (hadron) yields are provided as well.
The present results feature a larger data sample, an extended kinematic domain,
 and an improved treatment of the particle identification, when compared to the
results of Ref.~\citen{proc-nour}.
The 
$x$, $y$ and $z$ binning used in the analysis is given in 
Table~\ref{tab:Binning}. The $Q^2$ values range from $1\,(\GeV/c)^2$ at 
the smallest $x$ to about $60\,(\GeV/c)^2$ at the highest $x$, 
with $\langle Q^2 \rangle = 3\,(\GeV/c)^2$.

\renewcommand{\arraystretch}{1.1} 
 \begin{table}[!htbp]
  \caption{Bin limits for the three-dimensional binning in $x$, $y$ and $z$.}
  \label{tab:Binning}
 \centering
    \begin{tabular}{|l|rrrrrrrrrrrrr|}
      \hline
     &\multicolumn{13}{|c|} {bin limits}\\
      \hline
      $x$  &0.004 &0.01& 0.02&0.03&0.04&0.06&0.1&0.14&0.18&0.4&&&\\
     \hline
      $y$ & 0.1&0.15&0.2&0.3&0.5&0.7&&&&&&&\\
     \hline
      $z$ & 0.2&0.25&0.3&0.35&0.4& 0.45&0.5&0.55&0.6&0.65&0.7&0.75&0.85\\
      \hline
     \end{tabular}
\end{table}
 
\begin{figure}[htbp]
\centering
\includegraphics[width=\textwidth]{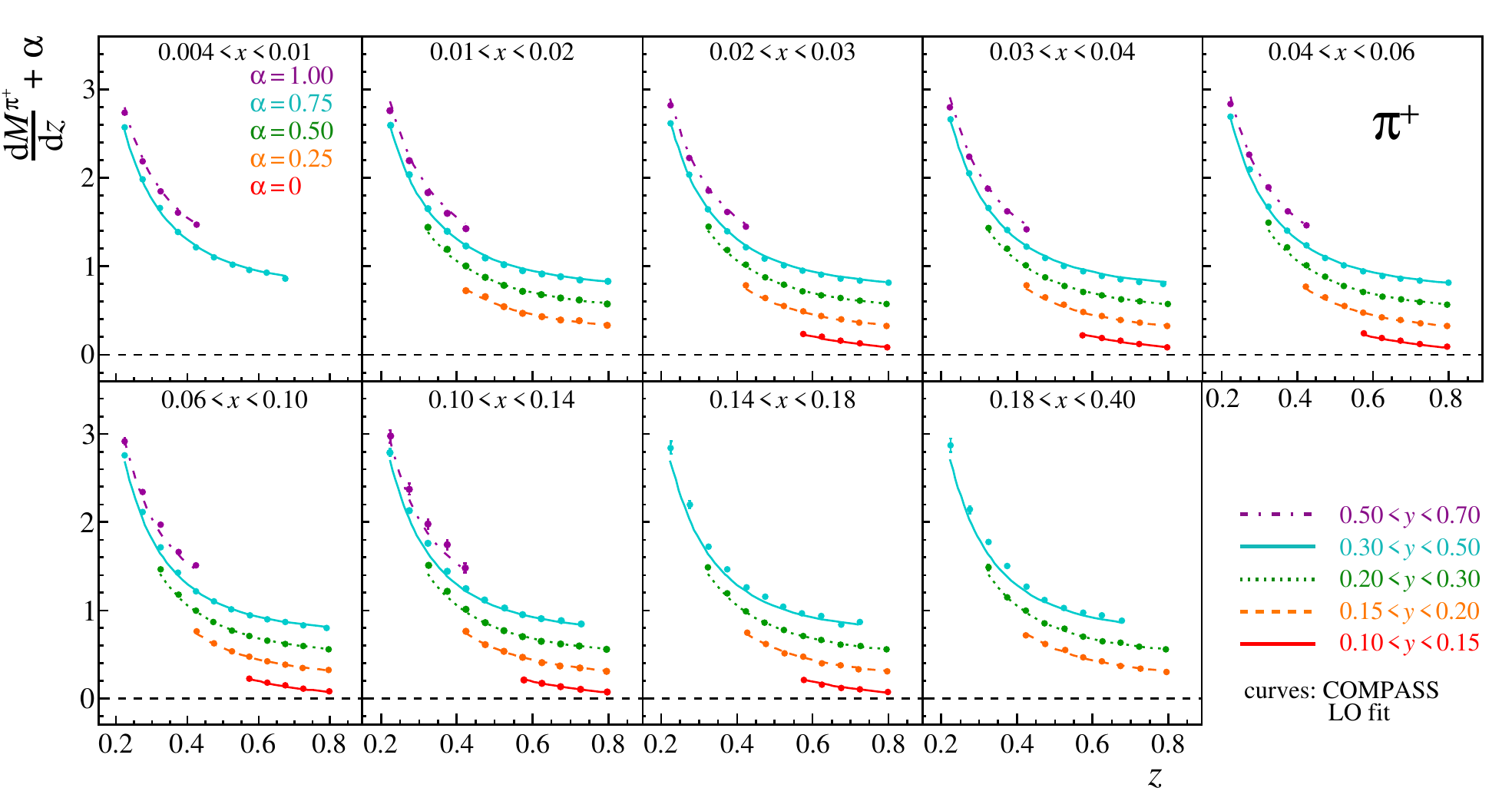}
\caption{Positive pion multiplicities versus $z$ for nine $x$ bins and five 
$y$ bins (for clarity staggered vertically by $\alpha$). Only statistical uncertainties are shown. 
The curves correspond to the COMPASS LO fit (see 
Section~\ref{sec:fragmentation}). (Coloured version online)}
\label{res1}

\includegraphics[width=\textwidth]{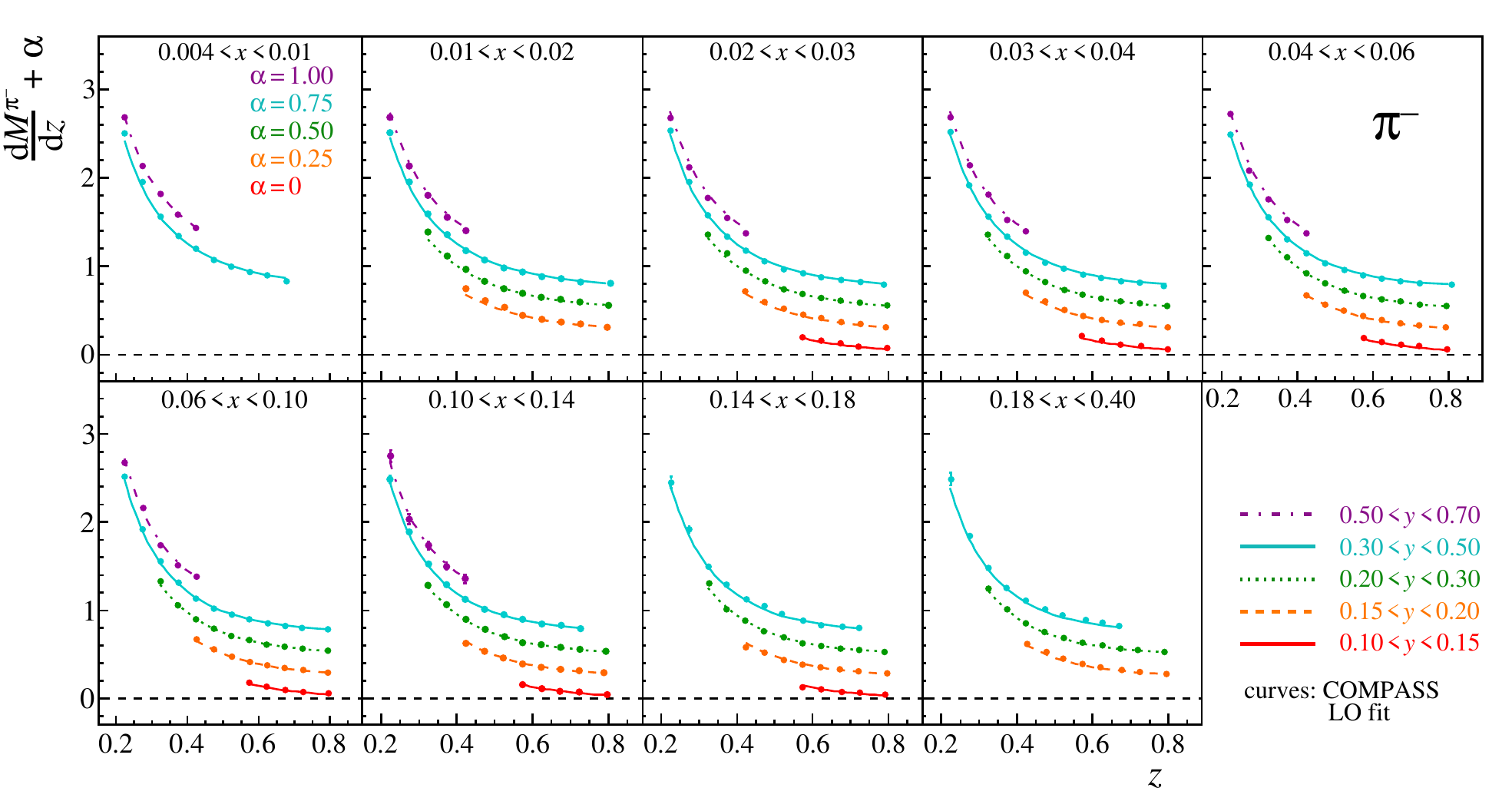}
\caption{Same as Fig.~\ref{res1} for negative pions. (Coloured version online)} 
\label{res2}
\end{figure}
In Figures~\ref{res1} and \ref{res2}, the results for the $z$ and $y$ dependences of the $\pi ^+$ and $\pi ^-$
multiplicities are presented in the nine bins of $x$. Only statistical uncertainties 
are shown, which are in most cases smaller than the size of the symbols. The curves 
correspond to the LO pQCD fit as discussed in Section~\ref{sec:fragmentation}. 

\begin{figure}[htbp]
\centering
\includegraphics[width=0.9\textwidth]{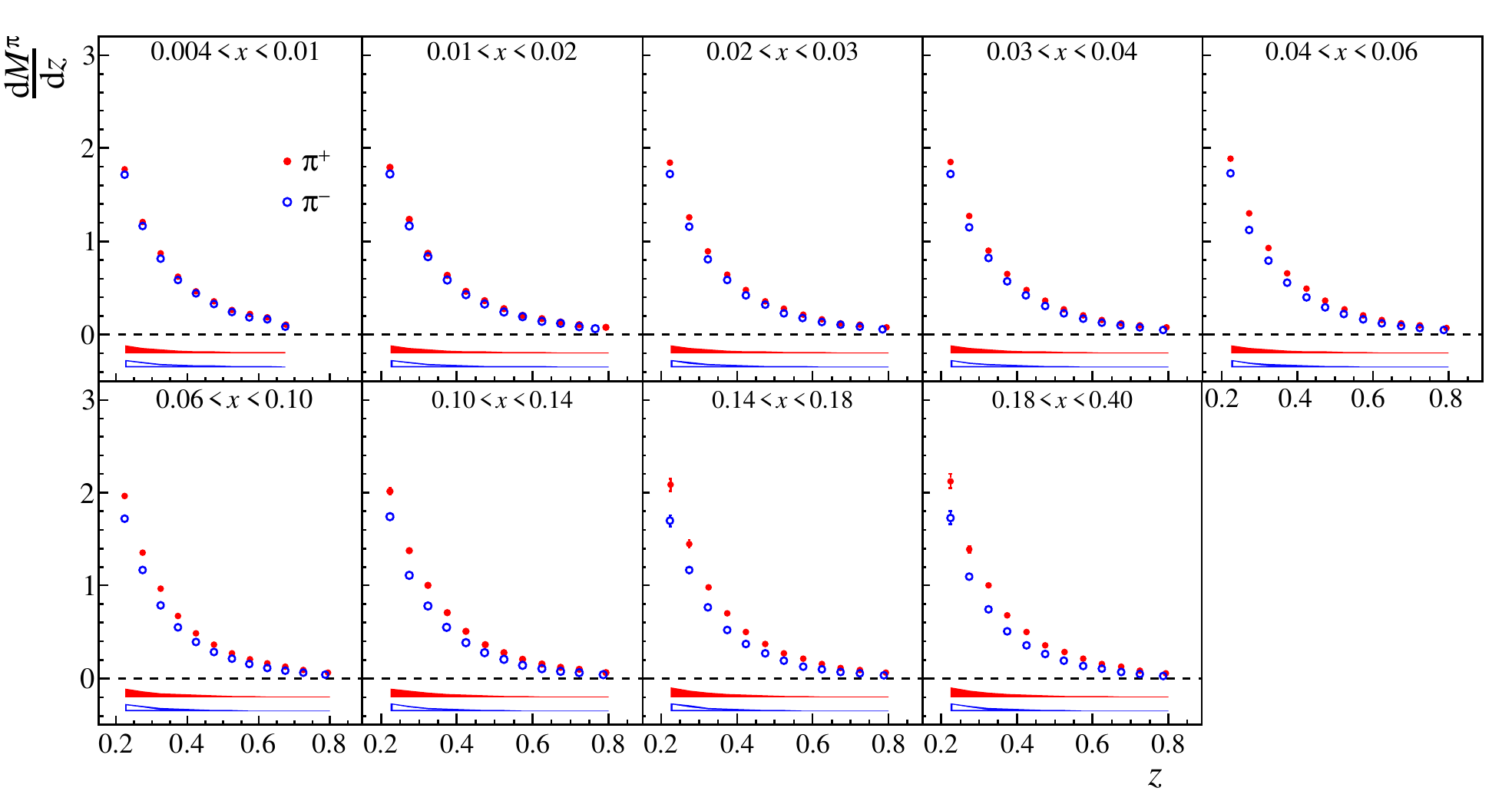}
 \caption{Positive (closed) and negative (open) pion multiplicities versus $z$ for 
nine $x$ bins. The bands correspond to the total systematic uncertainties.}
 \label{res3}

\includegraphics[width=0.9\textwidth]{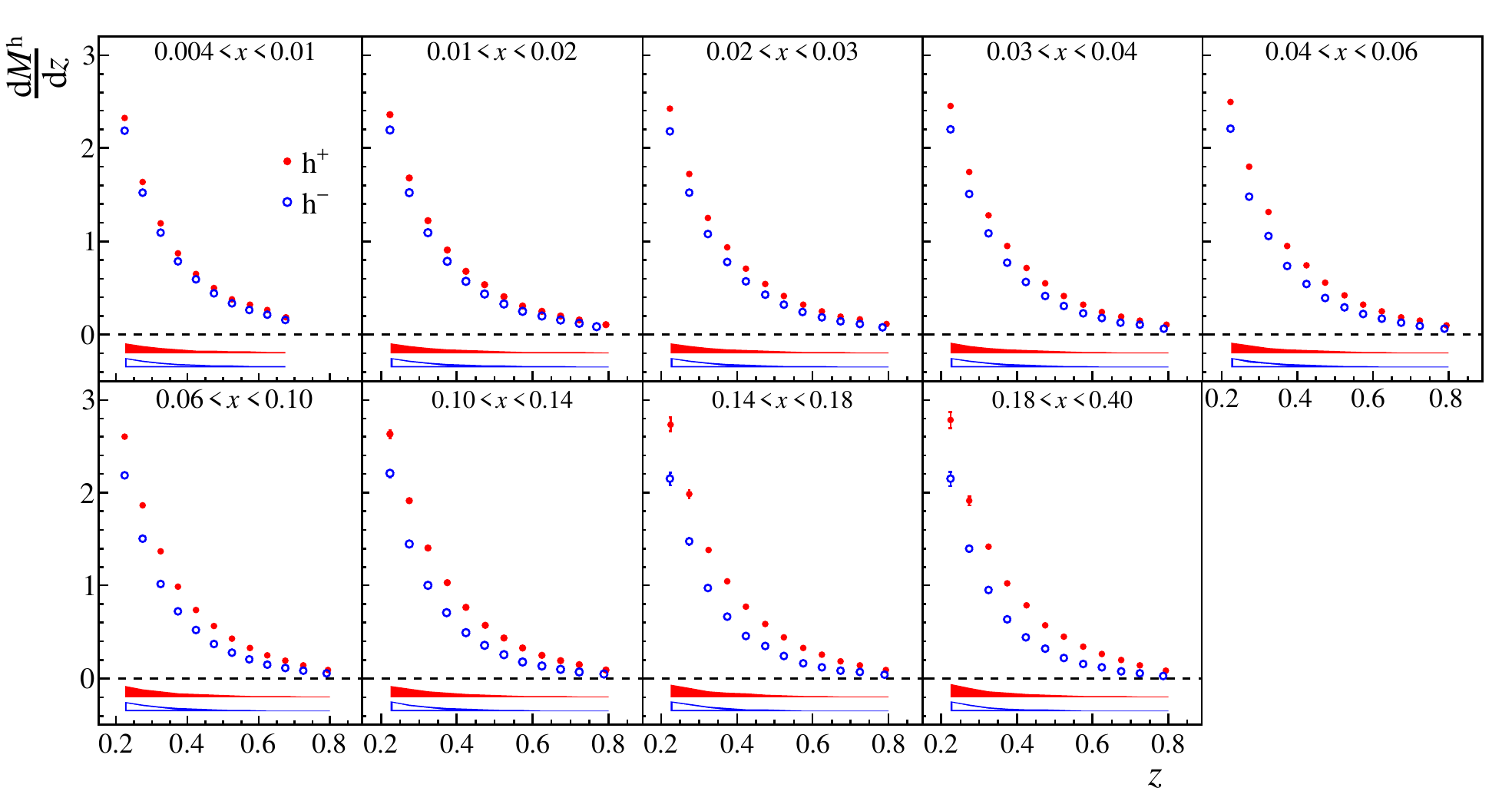}
\caption{Same as Fig.~\ref{res3} for charged unidentified hadrons.}
\label{res4}
\end{figure}
In Figure~\ref{res3} multiplicities of positively (closed circles) and 
negatively (open circles) charged pions are shown versus $z$, separately for 
the nine $x$ bins but averaged over $y$. The error bars correspond to the 
statistical uncertainties and the bands to the total systematic ones. 
Figure~\ref{res4} shows the results for the
charged hadron multiplicities. Both figures exhibit a strong dependence 
on $z$, as already observed in previous measurements, and a weak one on $x$.
Multiplicities are higher for positively than for negatively charged hadrons 
because of u-quark dominance. This difference is more pronounced for 
unidentified hadrons than for pions since negative kaons and antiprotons, 
which are included in the  hadron sample, do not contain nucleon valence 
quarks.

The sum of positively and negatively charged hadron multiplicities integrated over 
$z$ is of special interest. As noted in Ref.~\citen{HERMES2008}, in the 
case of kaon multiplicities the sum is used 
to extract at LO pQCD the product of the strange quark distribution and the 
fragmentation function of strange quarks into kaons. The summed  
$\pi^+$ and $\pi^-$ multiplicities allow us to verify the applicability 
of the LO  pQCD formalism in the COMPASS kinematic domain.
For an isoscalar target and taking into account only two independent quark FFs 
$D^{\pi}_{\fav}$ and $D^{\pi}_{\unf}$ (see Section~\ref{sec:fragmentation}), the sum of $\pi^{+}$ 
and $\pi^{-}$ multiplicities integrated over $z$ can be written at LO as
\begin{equation}
{\mathscr M}^{\pi^+}+{\mathscr M}^{\pi^-}
=  {\mathscr D}^{\pi}_{\fav} + {\mathscr D}^{\pi}_{\unf} - 
\frac{2S}{5U+2S} ({\mathscr D}^{\pi}_{\fav}-{\mathscr D}^{\pi}_{\unf}),
\end{equation}
with ${\mathscr M}^{\pi^{\pm}}= 
\int \langle M^{\pi^{\pm}}(x,y,z) \rangle_y\,{\text d}z$. The 
combinations of PDFs $U=u+\bar{u}+d+\bar{d}$ and $S=s+\bar{s}$ 
depend on $x$ and $Q^2$, and ${\mathscr D}^{\pi}_{\fav}(Q^2)=\int D^{\pi}_{\fav}(z,Q^2)\,{\text d}z $ 
and ${\mathscr D}^{\pi}_{\unf}(Q^2)=\int D^{\pi}_{\unf}(z,Q^2)\,{\text d}z $ are integrated 
over the measured $z$ range and depend on $Q^2$ only.
The pion multiplicity sum is expected to be almost flat in $x$, 
as the term $2S/(5U+2S)$ is small
and the $Q^2$ dependence of ${\mathscr D}^{\pi}_{\fav} + {\mathscr D}^{\pi}_{\unf}$ 
is rather weak (of the order of 3\%)~\cite{deFlorian:2007aj}.

\begin{figure}[htbp]
\centering
\includegraphics[width=0.49\textwidth]{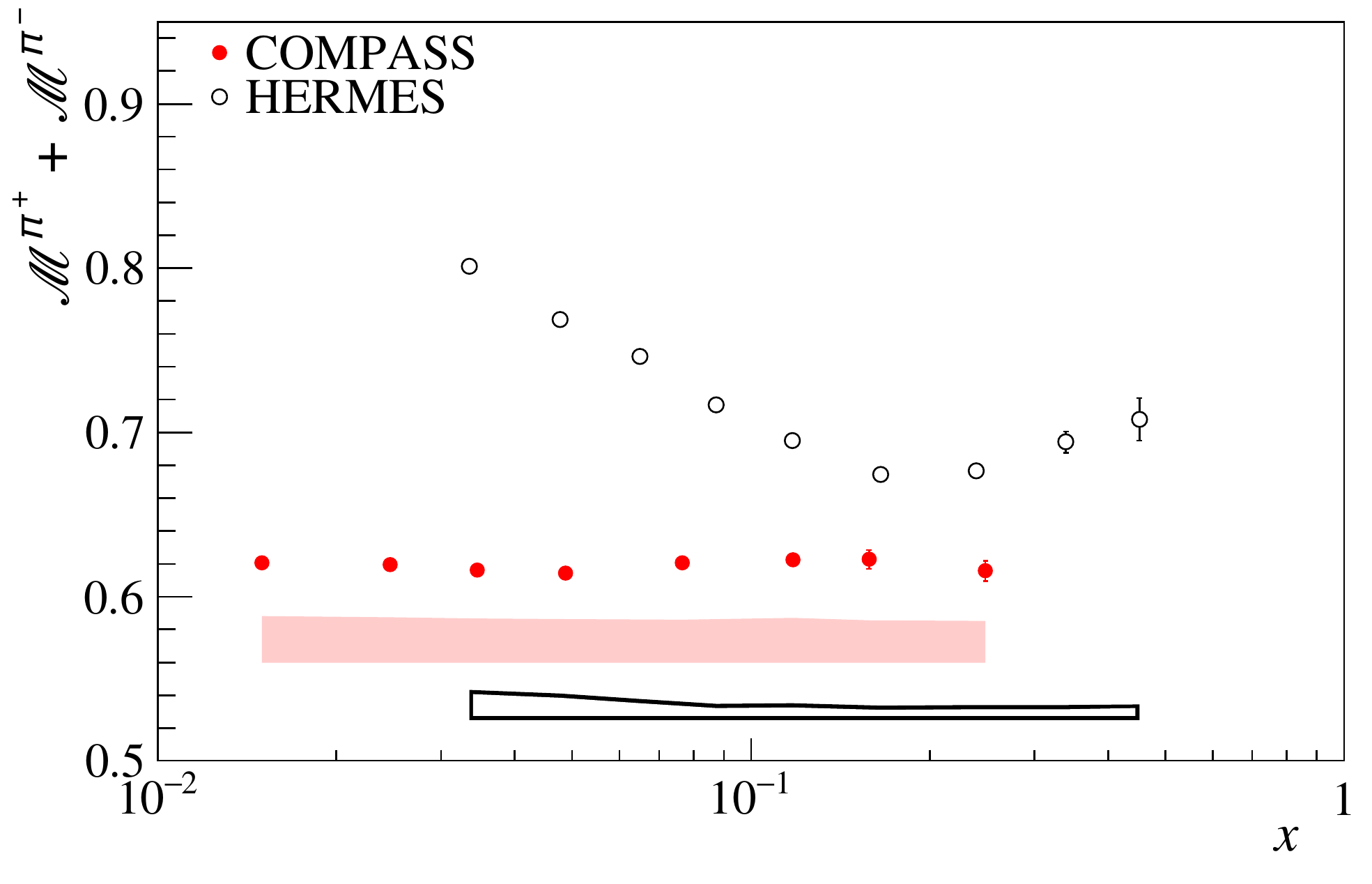}
\includegraphics[width=0.49\textwidth]{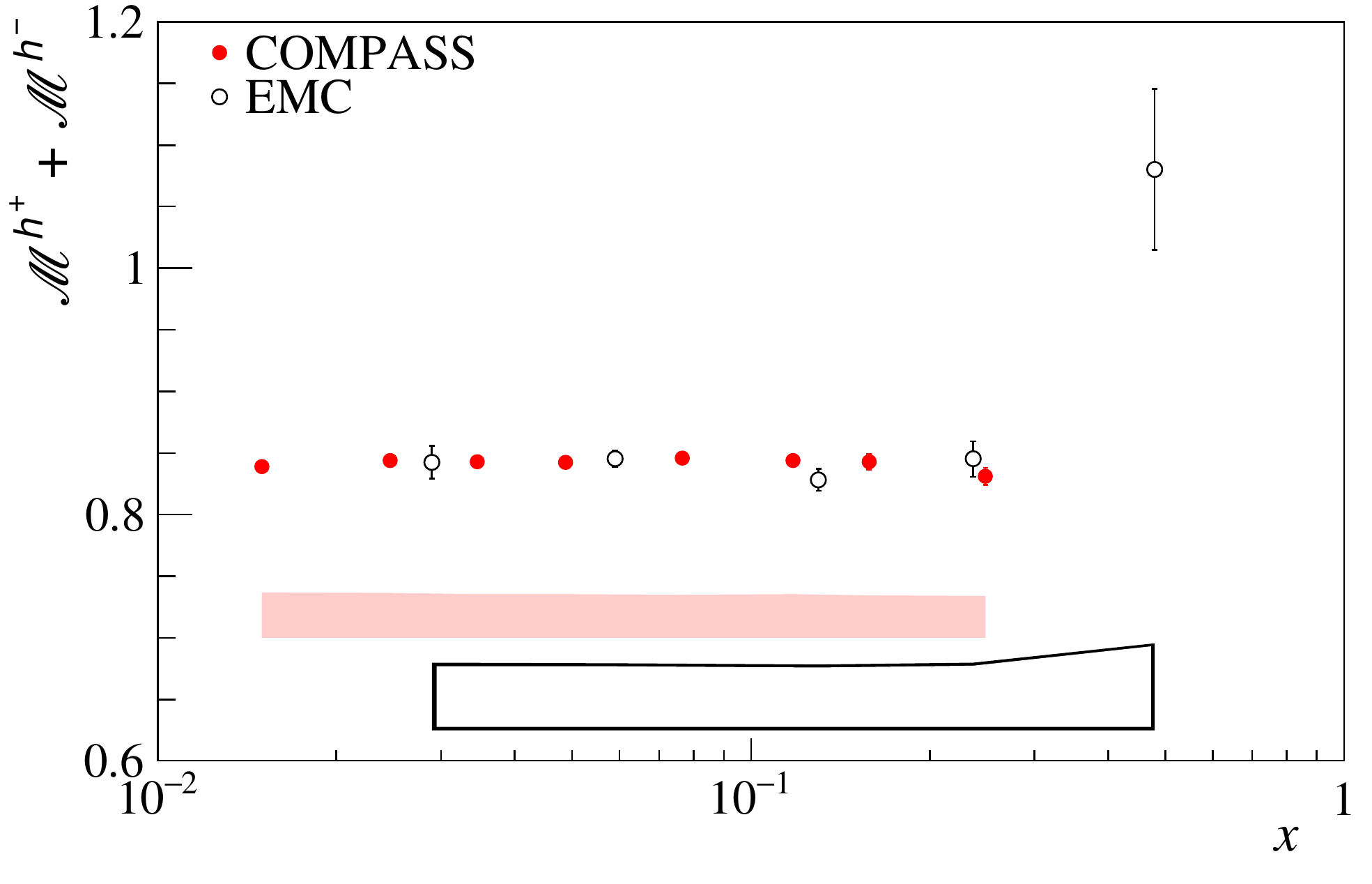}
\caption{Left: Sum of  
${\mathscr M}^{\pi^{+}}$ and ${\mathscr M}^{\pi^{-}}$ versus $x$. The COMPASS data (closed circles) are compared to HERMES results (open circles); 
Right: Sum of ${\mathscr M}^{{\rm h}^{+}}$ and 
${\mathscr M}^{{\rm h}^{-}}$ versus $x$. The
COMPASS data (closed circles) are compared to EMC results (open circles).  The 
systematic uncertainties are shown as bands at the bottom.}
\label{fig:Mult_sum} 
\end{figure}

Figure~\ref{fig:Mult_sum}\,(left) shows the result for the sum $\mathscr M^{\pi^+} + \mathscr M^{\pi^-}$
of $\pi^+$ and $\pi^-$ multiplicities, integrated 
over $z$ from 0.2 to 0.85 and averaged over $y$ between 0.1 and 0.7, 
as a function of $x$. 
The expected weak $x$ dependence is indeed observed in the 
data. 
In the same figure, the results of HERMES~\cite{HERMES} integrated over $z$ 
from 0.2 to 0.8 are shown using the so-called $x$ representation.
The HERMES multiplicities are larger and show a different dependence on $x$. 
Note however that the HERMES data were measured at a lower energy
and correspond to different kinematics.
In order to compare the COMPASS results also with the EMC ones \cite{EMC}, 
the sum of unidentified-charged-hadron multiplicities 
is shown in Fig.~\ref{fig:Mult_sum} (right). 
The results from COMPASS and EMC, which correspond to comparable
kinematics, are found in excellent agreement.

\begin{figure}[htbp]
\centering
\includegraphics[width=0.49\textwidth]{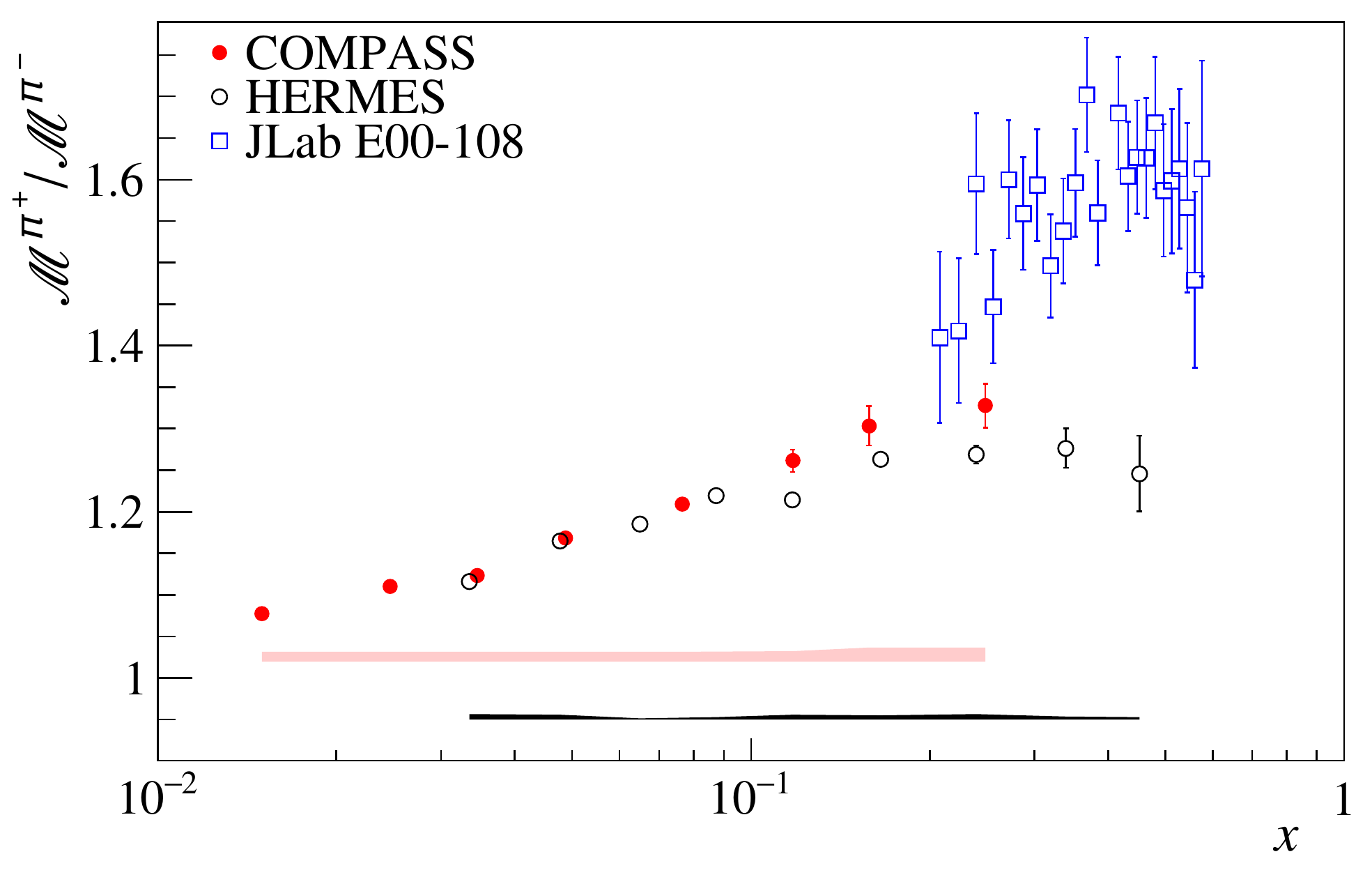}
\includegraphics[width=0.49\textwidth]{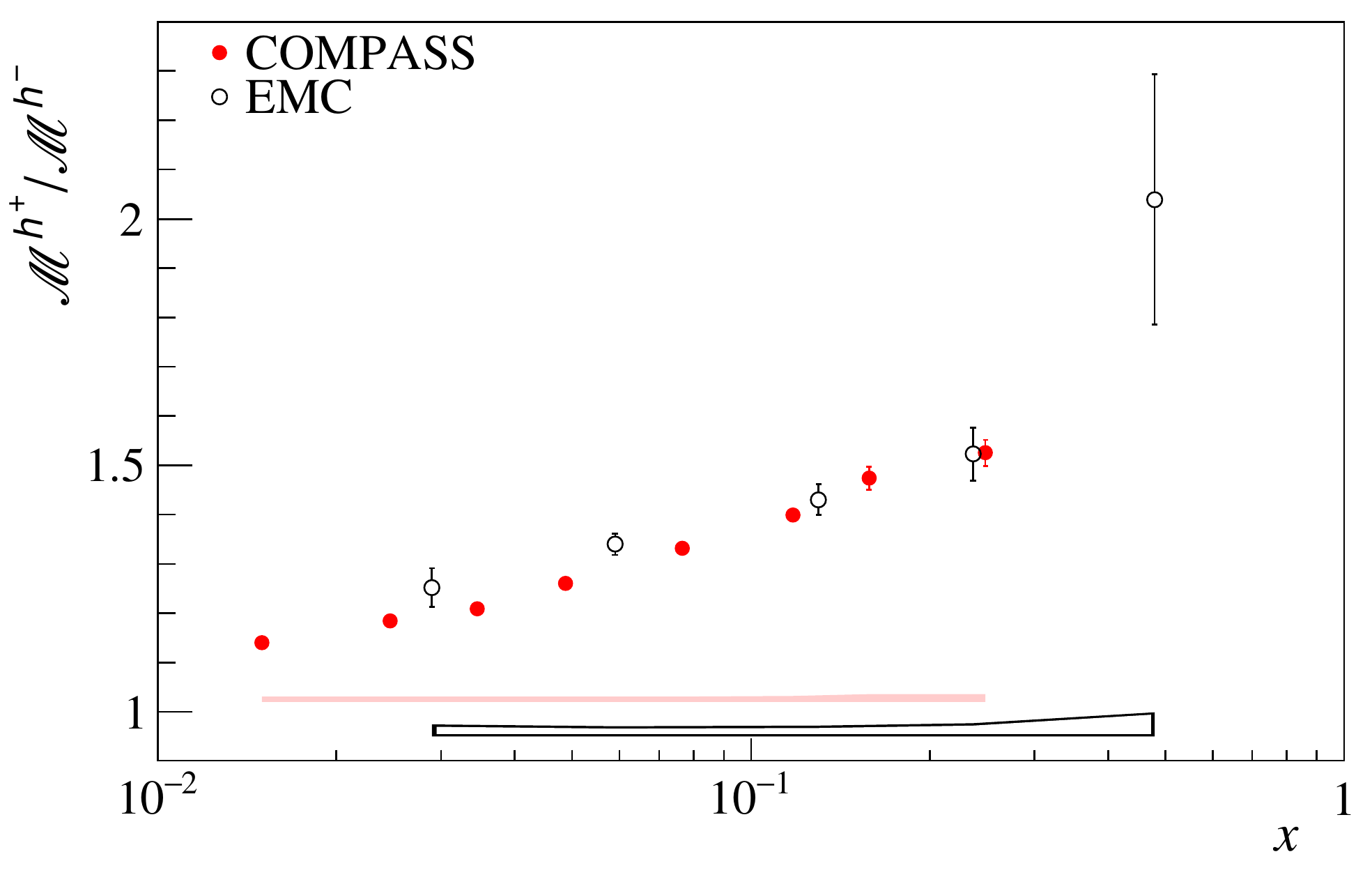}
\caption{Left: Ratio ${\mathscr M}^{\pi^{+}}/{\mathscr M}^{\pi^{-}}$ versus $x$ from COMPASS (closed points), HERMES (open circles) 
and JLab (open squares). Right: Ratio  
${\mathscr M}^{{\rm h}^{+}}/{\mathscr M}^{{\rm h}^{-}}$ versus $x$ for COMPASS (closed circles) and EMC (open circles) results.
The systematic uncertainties are shown as bands at the bottom.}
\label{fig:ratio}
\end{figure}

Another quantity of interest is the $x$ dependence of the ratio 
${\mathscr M}^{\pi^{+}}/{\mathscr M}^{\pi^{-}}$, where most experimental 
systematic effects cancel. The results are shown in Fig.~\ref{fig:ratio} 
(left) as a function of $x$. They are in reasonable agreement with the 
HERMES values in the measured range. The values 
obtained from the JLab E00-108 experiment~\cite{JLAB108} for $z>0.3$ at higher $x$ and lower $W$ 
values are also shown for completeness.
In Figure~\ref{fig:ratio} (right), the ratio ${\mathscr M}^{{\rm h}^{+}}/{\mathscr M}^{{\rm h}^{-}}$ calculated for unidentified hadron 
multiplicities is shown for COMPASS and EMC data. These results are in excellent 
agreement.


\section{Extraction of quark-to-pion fragmentation functions}
\label{sec:fragmentation}
The present data on pions cover a wide kinematic range in $x$ and $z$
and represent an important input for the extraction of quark-to-pion FFs
in future NLO pQCD analyses of the world data.
We present here an extraction of quark-to-pion FFs,
however restricted to the present pion data and limited to LO pQCD.
The results are
checked by performing an independent direct extraction of the two 
quark FFs in fixed kinematic bins.

The fragmentation of a quark of a given species into a final-state hadron 
is called favoured if the quark flavour corresponds to a valence quark in the hadron, otherwise
the fragmentation is called unfavoured. 
According to isospin and charge symmetry, and assuming in 
addition that the strange quark FF is equal to the other unfavoured FFs, only 
two independent quark-to-pion FFs remain:
\begin{eqnarray}
\Dfav^\pi = D_{\text{u}}^{\pi^+} = D_{\dbar}^{\pi^+} = D_{\text{d}}^{\pi^-} = D_{\ubar}^{\pi^-}
\nonumber\\
\Dunf^\pi= D_{\text{d}}^{\pi^+} =  D_{\ubar}^{\pi^+} = D_{\text{u}}^{\pi^-} =  D_{\dbar}^{\pi^-} = D_{\text{s}}^{\pi^\pm} = D_{\sbar}^{\pi^\pm}\,.
\label{eq:pionfragmentation}
\end{eqnarray}
A LO pQCD fit to the present set of $\pi^+$ and $\pi^{-}$ multiplicities
in ($x$, $y$, $z$) bins is performed to extract a parametrisation of these two 
FFs as a function of $z$.
For the evolution
to the $Q^2$ value of a given data point, the DGLAP $Q^2$ evolution code of 
Ref.~\citen{HiraiKumano} is used. Even at LO, this 
evolution involves the additional gluon FF, $\Dg^\pi = \Dg^{\pi^+}= \Dg^{\pi^-}$. 
For the PDFs, MSTW08 at 
LO~\cite{MSTW08} is used. 
The following functional form is assumed for 
the $z$ dependence of the FFs:

\begin{equation}
zD_i(z) = N_i\frac{z^{\alpha_i} (1-z)^{\beta_i}}{ \int_{0.2}^{0.85}  z^{\alpha_i} (1-z)^{\beta_i}\text{d}z},
\label{eq:complexParametrization}
\end{equation}

where the reference scale is $Q_0^2 = 1\,(\GeV/c)^2$ 
and $i=\{{\rm fav,unf,g}\}$.
In order to fit simultaneously $M^{\pi^{+}}$ and $M^{\pi^{-}}$, 
a $\chi ^2$ minimisation procedure is applied. It takes into account
the quadratic sum of the statistical and uncorrelated systematic 
uncertainties, $\Delta^2$:
\begin{equation}
\chi^2 = \sum_{\pi^+,\pi^-} \sum_{j=1}^N\left[\left(\frac{\sum_q e_q^2 q\left(x_j, Q_j^2\right) 
D_q^{\pi^\pm}\left(z_j, Q_j^2\right)}
{\sum_q e_q^2 q\left(x_j, Q_j^2\right)} - 
M_{\text{exp}}^{\pi^\pm}\left(x_j, Q_j^2,z_j\right)\right) / \Delta_j\right]^2,
\end{equation}
where $N$ denotes the number of data points.
The multiplicities are well described by the fit as shown in 
Figs.~\ref{res1} and ~\ref{res2}. 

\begin{figure}[!h]
\centering
\includegraphics[width=.49\textwidth]{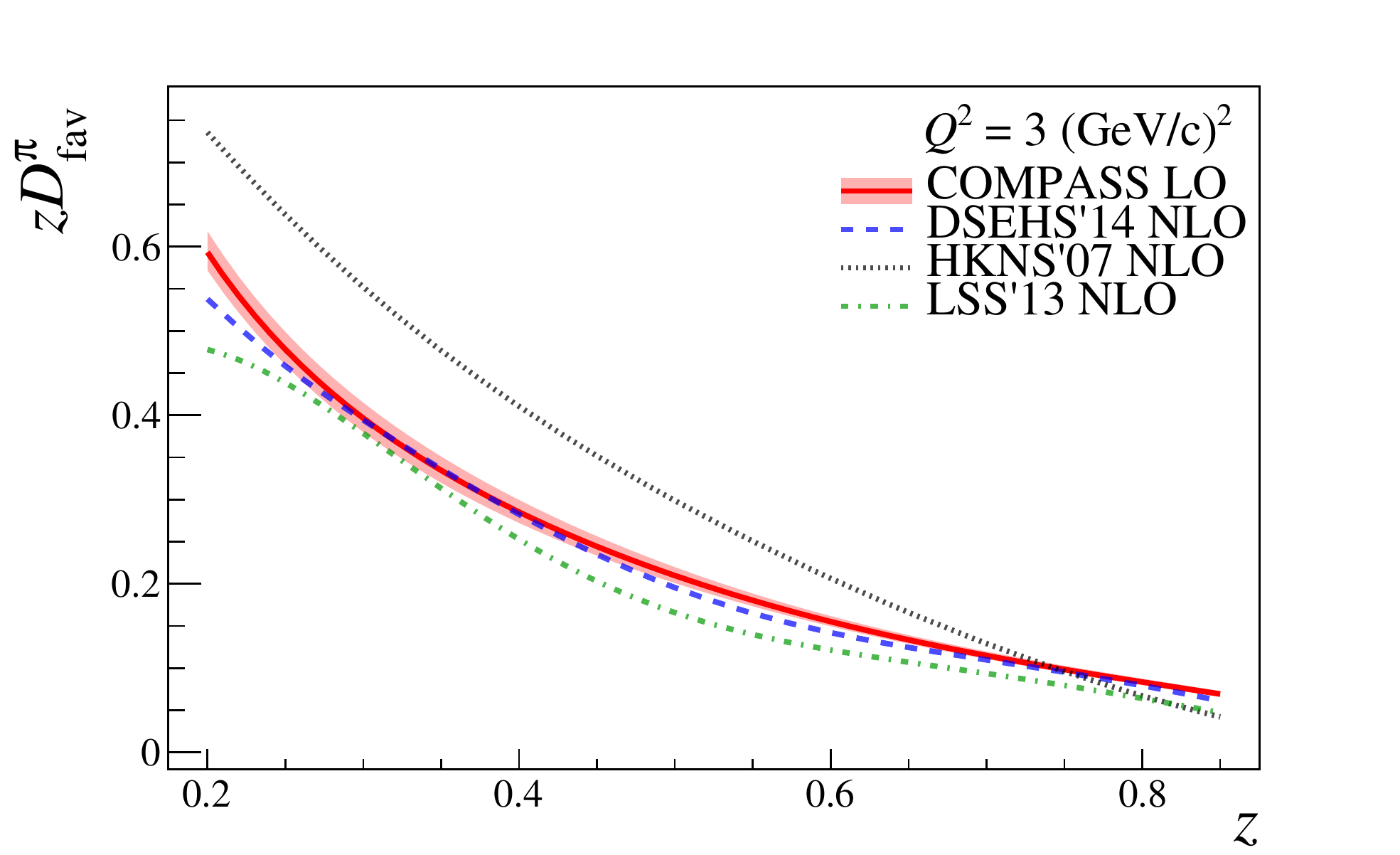}

\includegraphics[width=.49\textwidth]{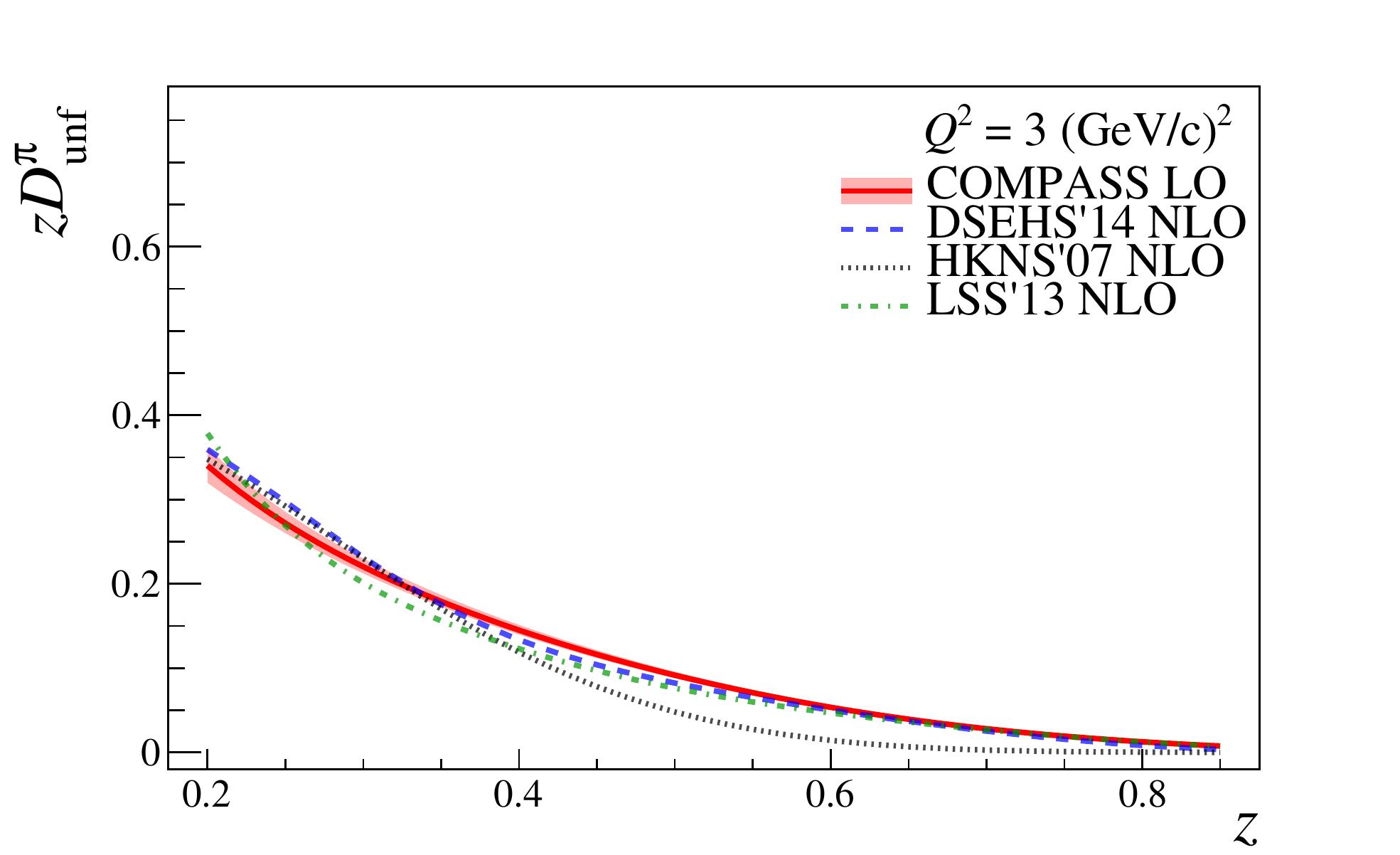}

\includegraphics[width=.49\textwidth]{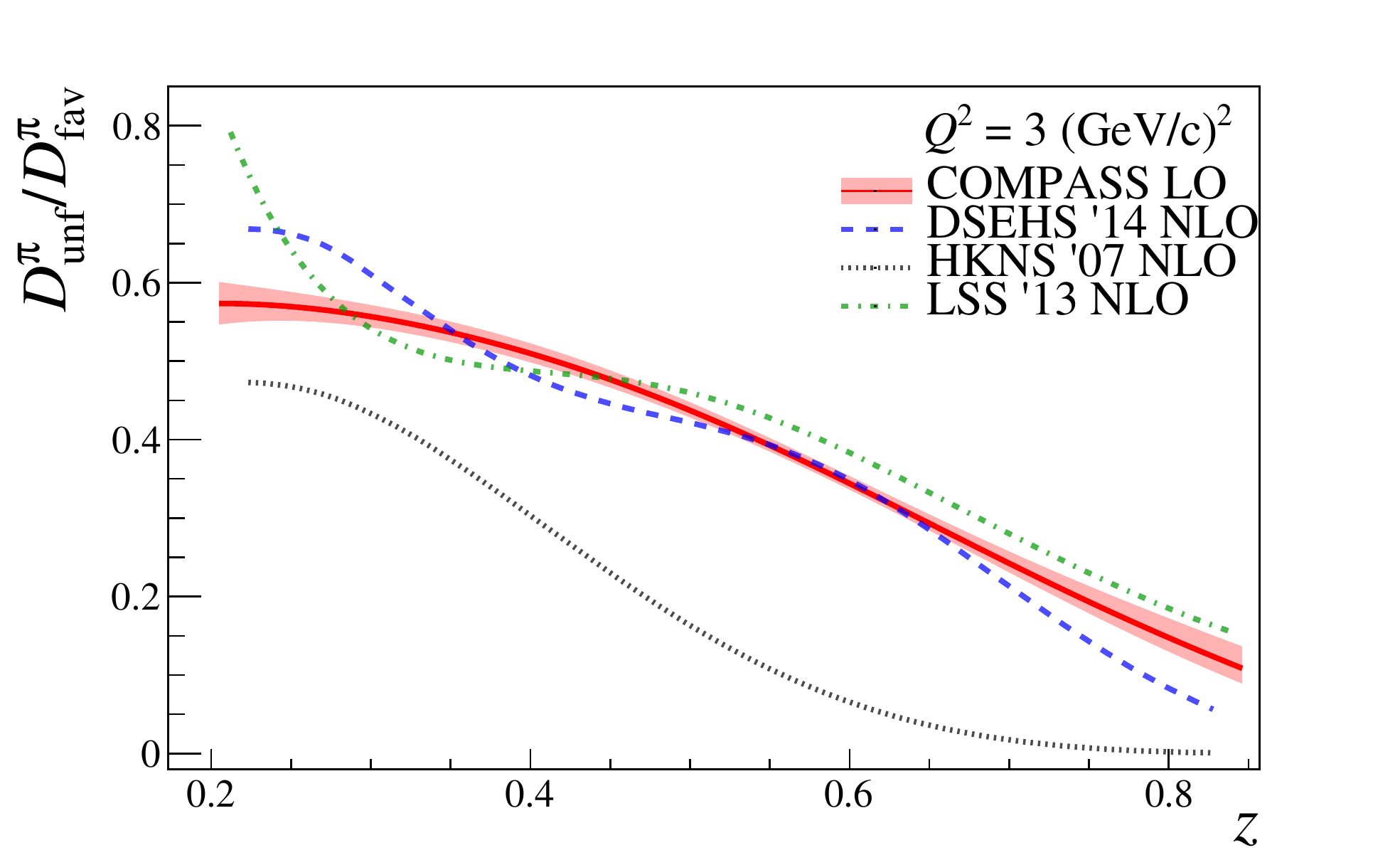}
\caption{Favoured (top) and unfavoured (middle) quark-to-pion FFs and the ratio 
$D^{\pi}_{\rm unf}/D^{\pi}_{\rm fav}$ (bottom), as obtained from the
COMPASS LO fit, compared to the DSEHS, HKNS and LSS fits at NLO.
The bands represent the total uncertainties for the FFs and the total statistical uncertainty for the ratio (see text). (Coloured version online)}
\label{fig:FFvsNLO}
\end{figure}
The results for the extracted favoured and 
unfavoured quark FFs and the ratio $D^{\pi}_{\rm unf}/D^{\pi}_{\rm fav}$
are shown in 
Fig.~\ref{fig:FFvsNLO} as a function of $z$ {evolved to} $Q^2=3\,(\GeV/c)^2$. The unfavoured FF 
is smaller than the favoured one, as expected,  and their ratio is seen to decrease with $z$.
The shaded bands depict the total uncertainty. These bands are determined 
using a MC sampling 
method~(bootstrap method in Ref.~\citen{NumericalRecipes}). One hundred 
replicas of the 
original data set are built by generating standard normal deviates as noise 
factors that 
are then multiplied by the data point uncertainties
and added to the original data point. For the statistical uncertainty and uncorrelated systematic 
uncertainty
the noise factor is generated for each point separately, while
for the correlated systematic uncertainty the noise factor is generated
only once per replica. The bands widths are given by the root mean square
of the corresponding distributions.
Note that for the ratio $D_{\rm fav}^{\pi}/D_{\rm unf}^{\pi}$
most of the correlated systematic uncertainties cancel.

In Figure~\ref{fig:FFvsNLO}, also recent parametrisations of FFs obtained from NLO 
analyses by the LSS~\cite{LSS-FF}, DSEHS~\cite{DSS15} and HKNS~\cite{HKNS-FF} groups are 
compared to the COMPASS LO fit.
While the present fit disagrees with the HKNS parametrisation based on
electron--positron annihilation data only, qualitative agreement is
obtained with the DSEHS and LSS parametrisations that include, in
addition to HERMES data, preliminary COMPASS data based on only a
fraction of the presently analysed data with a reduced kinematic
coverage and larger systematic uncertainties. Therefore, the impact of
the present COMPASS results on the global fits will be considerably
enhanced when comparing to the impact of the preliminary data
discussed in Ref.~\citen{proc-nour}.

\begin{figure}[!htp]
\centering
\includegraphics[width=0.6\textwidth]{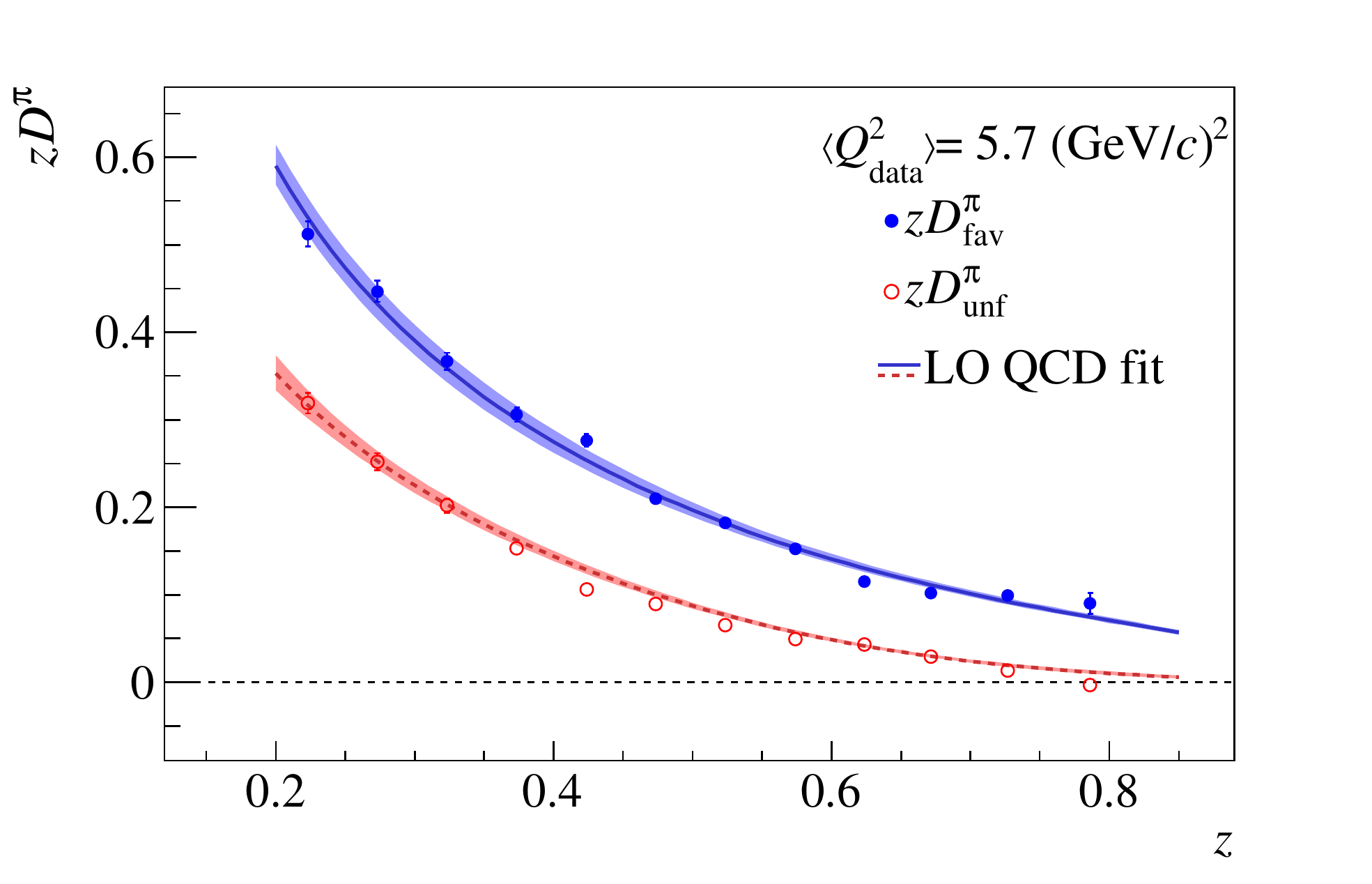}
\caption{The $z$ dependence of the favoured (closed symbols) and unfavoured (open 
symbols) quark-to-pion fragmentation functions,  
$zD_{\rm fav}^{\pi}$ and $zD_{\rm unf}^{\pi}$,
extracted directly from pion multiplicities 
for the bin $0.04<x<0.06$ and the bin $0.1<y<0.15$. For comparison, the result
from the present LO QCD fit to the pion multiplicities is shown at $Q^2=6\,(\GeV/c)^2$.  
The bands represent the total uncertainties of the QCD fit (see text). } 
\label{fig:ff2}
\end{figure}

An alternative method to extract the FFs from the pion multiplicity data is to solve
the system of two linear equations for $M^{\pi^+}$ and $M^{\pi^-}$ in each 
$(x,y,z)$ bin for the values 
of $D^{\pi}_{\fav}(\langle z \rangle,\langle Q^2 \rangle)$ 
and $D^{\pi}_{\unf}(\langle z \rangle,\langle Q^2 \rangle)$  by using 
Eqs.\,(\ref{MulDef}) and (\ref{Sigma}). 
No functional form has to be assumed, and the DGLAP evolution 
for FFs is not needed. The same PDFs as mentioned above are used.
The results from this direct extraction of FFs are in good agreement with the 
results obtained from the LO fit. This is illustrated in Fig.~\ref{fig:ff2} for
one ($x$, $y$) bin.

\section{Summary and conclusions}
We have presented differential multiplicities of charge-separated pions and unidentified charged
hadrons measured in SIDIS of muons off an isoscalar target. {The results are given in 
3-dimensional} bins of $x$, $y$ and $z$ and cover the kinematic range $Q^2>1\,(\GeV/c)^2$, 
$0.004  <x<0.4$ and 
$0.2<z<0.85$. The numerical values  are available in Ref.~\citen{HEPDATA}
with and without the subtraction of the contribution of diffractive 
vector meson production to SIDIS. In addition the radiative corrections
factors are provided. These high precision 
multi-dimensional data provide an important input
for future NLO QCD fits of fragmentation functions.

The sum of the $z$-integrated positive and negative unidentified hadron 
and pion multiplicities shows a flat $x$ behavior, as expected in LO pQCD. 
For unidentified hadrons this sum is
in agreement with EMC results taken at comparable kinematics,
whereas some inconsistency is observed when comparing to HERMES pion
data that were taken at different kinematics.
The ratio of the $z$-integrated positive and negative hadron and pion
multiplicities as a function of $x$ nicely confirms
the previous measurements from HERMES and EMC.

The measured charged pion multiplicities were used for a LO extraction of
the favoured and unfavoured pion FFs. 
While both FFs are significantly different from those obtained in the 
HKNS fit to only the
electron--positron annihilation data, they are in good agreement with those 
obtained in recent NLO
fits that also include a preliminary release of the present data.

\section*{Acknowledgements}
We gratefully acknowledge the support of the CERN management and staff and the
skill and effort of the technicians of our collaborating institutes.  This work
was made possible by the financial support of our funding agencies.  Special
thanks go to M.~Hirai and S.~Kumano for providing us with the DGLAP evolution 
code for fragmentation functions.

\bibliographystyle{prsty}

\input bibliography.tex
\end{document}

%% file: Authors2016.tex
%
%
\section*{The COMPASS Collaboration}
\label{app:collab}
\renewcommand\labelenumi{\textsuperscript{\theenumi}~}
\renewcommand\theenumi{\arabic{enumi}}
\begin{flushleft}
C.~Adolph\Irefn{erlangen},
J.~Agarwala\Irefn{calcutta},
M.~Aghasyan\Irefn{triest_i},
R.~Akhunzyanov\Irefn{dubna}, 
M.G.~Alexeev\Irefn{turin_u},
G.D.~Alexeev\Irefn{dubna}, 
A.~Amoroso\Irefnn{turin_u}{turin_i},
V.~Andrieux\Irefn{saclay},
N.V.~Anfimov\Irefn{dubna}, 
V.~Anosov\Irefn{dubna}
% ...
\end{flushleft}
%
%
\begin{Authlist}
\item \Idef{turin_p}{University of Eastern Piedmont, 15100 Alessandria, Italy}
\item \Idef{aveiro}{University of Aveiro, Department of Physics, 3810-193 Aveiro, Portugal} 
\item \Idef{bochum}{Universit\"at Bochum, Institut f\"ur Experimentalphysik, 44780 Bochum, Germany\Arefs{l}\Arefs{s}}
\item \Idef{bonniskp}{Universit\"at Bonn, Helmholtz-Institut f\"ur  Strahlen- und Kernphysik, 53115 Bonn, Germany\Arefs{l}}
\item \Idef{bonnpi}{Universit\"at Bonn, Physikalisches Institut, 53115 Bonn, Germany\Arefs{l}}
\item \Idef{brno}{Institute of Scientific Instruments, AS CR, 61264 Brno, Czech Republic\Arefs{m}}
\item \Idef{calcutta}{Matrivani Institute of Experimental Research \& Education, Calcutta-700 030, India\Arefs{n}}
\item \Idef{dubna}{Joint Institute for Nuclear Research, 141980 Dubna, Moscow region, Russia\Arefs{o}}
\item \Idef{erlangen}{Universit\"at Erlangen--N\"urnberg, Physikalisches Institut, 91054 Erlangen, Germany\Arefs{l}}
\item \Idef{freiburg}{Universit\"at Freiburg, Physikalisches Institut, 79104 Freiburg, Germany\Arefs{l}\Arefs{s}}
\item \Idef{cern}{CERN, 1211 Geneva 23, Switzerland}
\item \Idef{liberec}{Technical University in Liberec, 46117 Liberec, Czech Republic\Arefs{m}}
\item \Idef{lisbon}{LIP, 1000-149 Lisbon, Portugal\Arefs{p}}
\item \Idef{mainz}{Universit\"at Mainz, Institut f\"ur Kernphysik, 55099 Mainz, Germany\Arefs{l}}
\item \Idef{miyazaki}{University of Miyazaki, Miyazaki 889-2192, Japan\Arefs{q}}
\item \Idef{moscowlpi}{Lebedev Physical Institute, 119991 Moscow, Russia}
\item \Idef{munichtu}{Technische Universit\"at M\"unchen, Physik Department, 85748 Garching, Germany\Arefs{l}\Arefs{r}}
\item \Idef{nagoya}{Nagoya University, 464 Nagoya, Japan\Arefs{q}}
\item \Idef{praguecu}{Charles University in Prague, Faculty of Mathematics and Physics, 18000 Prague, Czech Republic\Arefs{m}}
\item \Idef{praguectu}{Czech Technical University in Prague, 16636 Prague, Czech Republic\Arefs{m}}
\item \Idef{protvino}{State Scientific Center Institute for High Energy Physics of National Research Center `Kurchatov Institute', 142281 Protvino, Russia}
\item \Idef{saclay}{CEA IRFU/SPhN Saclay, 91191 Gif-sur-Yvette, France\Arefs{s}}
\item \Idef{taipei}{Academia Sinica, Institute of Physics, Taipei 11529, Taiwan}
\item \Idef{telaviv}{Tel Aviv University, School of Physics and Astronomy, 69978 Tel Aviv, Israel\Arefs{t}}
\item \Idef{triest_u}{University of Trieste, Department of Physics, 34127 Trieste, Italy}
\item \Idef{triest_i}{Trieste Section of INFN, 34127 Trieste, Italy}
\item \Idef{triest_ictp}{Abdus Salam ICTP, 34151 Trieste, Italy}
\item \Idef{turin_u}{University of Turin, Department of Physics, 10125 Turin, Italy}
\item \Idef{turin_i}{Torino Section of INFN, 10125 Turin, Italy}
\item \Idef{illinois}{University of Illinois at Urbana-Champaign, Department of Physics, Urbana, IL 61801-3080, USA}   
\item \Idef{warsaw}{National Centre for Nuclear Research, 00-681 Warsaw, Poland\Arefs{u} }
\item \Idef{warsawu}{University of Warsaw, Faculty of Physics, 02-093 Warsaw, Poland\Arefs{u} }
\item \Idef{warsawtu}{Warsaw University of Technology, Institute of Radioelectronics, 00-665 Warsaw, Poland\Arefs{u} }
\item \Idef{yamagata}{Yamagata University, Yamagata 992-8510, Japan\Arefs{q} }
\end{Authlist}
%
%
\renewcommand\theenumi{\alph{enumi}}
\begin{Authlist}
\item [{\makebox[2mm][l]{\textsuperscript{*}}}] Deceased
\item \Adef{a}{Also at Instituto Superior T\'ecnico, Universidade de Lisboa, Lisbon, Portugal}
\item \Adef{b}{Also at Department of Physics, Pusan National University, Busan 609-735, Republic of Korea and at Physics Department, Brookhaven National Laboratory, Upton, NY 11973, USA}
\item \Adef{r}{Supported by the DFG cluster of excellence `Origin and Structure of the Universe' (www.universe-cluster.de)}
\item \Adef{d}{Also at Chubu University, Kasugai, Aichi 487-8501, Japan\Arefs{q}}
\item \Adef{e}{Also at KEK, 1-1 Oho, Tsukuba, Ibaraki 305-0801, Japan}
\item \Adef{g}{Also at Moscow Institute of Physics and Technology, Moscow Region, 141700, Russia}
\item \Adef{v}{Supported by Presidential grant NSh--999.2014.2}
\item \Adef{h}{Present address: RWTH Aachen University, III.\ Physikalisches Institut, 52056 Aachen, Germany}
\item \Adef{i}{Present address: Uppsala University, Box 516, 75120 Uppsala, Sweden}
\item \Adef{c}{Supported by the DFG Research Training Group Programme 1102  ``Physics at Hadron Accelerators''}
%
%
\item \Adef{l}{Supported by the German Bundesministerium f\"ur Bildung und Forschung}
\item \Adef{s}{Supported by EU FP7 (HadronPhysics3, Grant Agreement number 283286)}
\item \Adef{m}{Supported by Czech Republic MEYS Grant LG13031}
\item \Adef{n}{Supported by SAIL (CSR), Govt.\ of India}
\item \Adef{o}{Supported by CERN-RFBR Grant 12-02-91500}
\item \Adef{p}{\raggedright Supported by the Portuguese FCT - Funda\c{c}\~{a}o para a Ci\^{e}ncia e Tecnologia, COMPETE and QREN,
 Grants CERN/FP 109323/2009, 116376/2010, 123600/2011 and CERN/FIS-NUC/0017/2015}
\item \Adef{q}{Supported by the MEXT and the JSPS under the Grants No.18002006, No.20540299 and No.18540281; Daiko Foundation and Yamada Foundation}
\item \Adef{t}{Supported by the Israel Academy of Sciences and Humanities}
\item \Adef{u}{Supported by the Polish NCN Grant 2015/18/M/ST2/00550}
\end{Authlist}